\documentstyle[aps,prl]{revtex} 

\begin{document}



\wideabs{
\title { Spin dynamics and magnetic correlation
length in two-dimensional quantum Heisenberg antiferromagnets }


\bigskip


\author {P.Carretta$^{1}$, T.Ciabattoni$^{1}$, A.Cuccoli $^{2}$
E.Mognaschi$^{1}$, A.Rigamonti$^{1}$, V.Tognetti$^{2}$, and
P.Verrucchi$^{2}$.}




\address        {$^1$ Dipartimento di Fisica "A. Volta" and Unit\`a INFM di
Pavia, Via Bassi, 6 
                I-27100 Pavia, Italy. \\
                 $^2$ Dipartimento di Fisica dell'Universit{\`a} di Firenze 
and Unit\`a INFM di Firenze,
         Largo E. Fermi, 2  I-50125 Firenze, Italy. \\}


\date{\today}


\maketitle
\begin{abstract}
  The correlated spin dynamics and the temperature dependence of the
  correlation length $\xi(T)$ in two-dimensional quantum
  ($S=1/2$) Heisenberg antiferromagnets (2DQHAF) on  square lattice are
  discussed in the light of experimental results of proton spin
  lattice relaxation in  copper formiate tetradeuterate 
  (CFTD). In this compound the exchange constant is much smaller than
  the one in recently studied 2DQHAF, such as La$_2$CuO$_4$
  and Sr$_2$CuO$_2$Cl$_2$. Thus the spin dynamics can be probed in
  detail over a wider temperature range. The NMR relaxation rates turn
  out in excellent agreement with a theoretical mode-coupling
  calculation. The deduced temperature behavior of $\xi(T)$ is in
  agreement with high-temperature expansions, quantum Monte Carlo
  simulations and the pure quantum self-consistent harmonic
  approximation. Contrary to the predictions of the theories 
  based on the  Non-Linear $\sigma$ Model, no evidence of crossover between
  different quantum regimes is observed.
\end{abstract}
\pacs {PACS numbers: 76.60.Es, 75.40.Gb, 75.10.Jm}
}




In recent years, it has been argued that the pairing mechanism leading
to high-temperature superconductivity can occur in the presence of
quantum critical fluctuations, as the ones expected in two-dimensional
quantum ($S=1/2$) Heisenberg antiferromagnets (2DQHAF) \cite{A,B}.  
This early observation has triggered a novel interest on the
correlated spin dynamics in 2DQHAF. Several studies have been
carried out on the behavior of the magnetic correlation length $\xi$
as a function of temperature and/or disorder, and on the possibility
of driving quantum phase transitions by hole doping. 
In particular, for a given degree of
disorder (i.e. doping) the temperature is expected to cause the transition
to a regime with
peculiar quantum spin excitations, which could be responsible for novel
pairing mechanisms
\cite{Pines}.
In spite of an enormous amount of experimental and theoretical efforts this
issue is still
controversial \cite{KimLT97,Beard}. Thus, it is believed that it is
necessary first
to unravel the problem of the
spin dynamics and of the temperature dependence of $\xi(T)$ in pure 2DQHAF.


The 2DQHAF with nearest-neighbor
interaction on  square lattice, is described by the
Hamiltonian
\begin{equation}
{\cal H}=\frac J2\sum_{\bf j,a}\,{\bf S}_{\bf j}\cdot{\bf S}_{\bf
j+a} \,,\label{A}
\end{equation}
with $J>0$ and ${\bf a}=(0,\pm a),(\pm a,0)$, $a$ being the lattice constant. 
The 2DQHAF has been commonly described in terms of the
$2$-dimensional Quantum Non-Linear $\sigma$ Model
(QNL$\sigma$M) \cite{C}, with the action given by
\begin{equation}
S=\frac 1{2g}\,\int_{-\infty}^\infty\int_0^{u}\,d{\bf
x}d\tau\,(|\nabla{\bf n}|^2-|\partial_\tau{\bf
n}|^2)\,;\;\;\;|{\bf n}|^2=1\,,\label{NLSM}
\end{equation}
where ${\bf n(x)}$ is a unitary 3D vector field,
$g=c_s\Lambda/\rho_s$ and $u=c_s\Lambda/T$ are the
coupling and the imaginary-time cut-off respectively.
This mapping is rigorous only in the
large-spin limit, where $\rho_s$ and $c_s$ are the spin stiffness
and the spin-wave velocity of the correspondent magnetic model.
This approach can be extended to $S=1/2$ provided that $c_s$ and $\rho_s$ are
introduced as fitting
parameters. Then a rich and interesting  
phase diagram can be devised \cite{B,D,E}.
Accordingly, the  QNL$\sigma$M has arisen to a great popularity, both for basic
quantum magnetism studies as well as a good candidate to
describe the modifications in the microscopic properties of a 2DQHAF when it
becomes a high-$T_c$ superconductor
upon charge doping. 


One of the most striking predictions is the presence
of the so called \emph{quantum criticality}, in a well defined
temperature window. Below a finite value of $g$,
by increasing the temperature, the system should cross from the
\emph{renormalized classical} to the \emph{quantum critical} regime and
eventually to the classical behavior \cite{B,E}.


It was suggested that this scenario could be probed by studying the
temperature behavior of the static correlation length $\xi(T)$. The latter
can be directly 
estimated by neutron scattering \cite{F,H,I} or by the longitudinal NMR
relaxation rate,
which  provides an indirect measure of $\xi$ \cite{G,GG}.
While for small $\xi$ ($\simeq a$), neutron scattering studies become difficult
and accuracy problems arise, it has been proved \cite{G} that NMR
relaxation, driven by the correlated spin dynamics, can be a
suitable tool to derive the temperature and doping dependence of the 
magnetic correlation length.



The behavior of $\xi$, according to the QNL$\sigma$M
approach, considerably changes in the various regimes\cite{B,E,F}.
Until now, most of the discussion on the correlation length in 2DQHAF
with $S=1/2$ has dealt with La$_2$CuO$_4$ and Sr$_2$CuO$_2$Cl$_2$.
However, in these compounds the large value of $J\sim1500\, K$
prevents one to obtain experimental results above $T/J\sim 0.6$ and
thus a detailed comparison with the theoretical predictions is
difficult. As a matter of fact the region where the quantum critical
regime is expected, according to the QNL$\sigma$M approach, should be
$0.4<T/J< 0.8$.  Neutron scattering experiments in
antiferromagnets with larger $S$ have been recently carried out with
good accuracy \cite{H,I} and no evidence of crossover has been found
in a large range of $T/J$. Indeed these data are in agreement with
high-temperature expansion (HTE) results \cite{J}, quantum Monte Carlo
(QMC) data \cite{K} and with the results of the pure quantum
self-consistent harmonic approximation (PQSCHA) recently obtained by
some of us \cite{reviewJPCM,L}. The latter approach is reliable for any
realistic temperature range for $S\geq 1$, and it gives good
results starting from $T/J\gtrsim0.35$ for $S=1/2$ \cite{L}.


The advantage of CFTD is that besides being a good
prototype of $S=1/2$ 2DQHAF on the square lattice (the
interlayer interaction constant is less than $10^{-4} J$), still
$J$ is rather small $(J\sim 80\,K)$ so that the experiments
can be carried out in a wide range of $T/J$. Thus the
classical behavior is approached and  a comprehensive
analysis becomes possible. In particular, one can explore
also the temperature range where the correlation length is of the same 
order  of the lattice constants $a$. 
$^1$H NMR experiments have been
performed on a $5\times 5\times 4$ mm$^3$ single crystal of copper formiate
tetradeuterate (Cu(HCO$_2$)$_2$.4D$_2$O) \cite{Okada}. The spin-lattice
relaxation rate $1/T_1$ has been measured in a $8$ kGauss
magnetic field ($g\mu_B H\ll J$) applied perpendicular to
the $ab$ plane, by using standard pulse sequences. In Fig.1
the experimental data for $1/T_1$ are reported.



The relaxation rates can be calculated, through the mode-coupling (MC)
approach \cite{Hubbard71,Kawasaki76,MCOU}. First, we calculate the
normalized Kubo relaxation function\cite{Lovesey}:
\begin{equation}
F_{\bf k}(t)= \frac {R_{\bf k}(t)}{R_{\bf k}(0)},\; \hspace{20pt}  {R_{\bf
k}(t)}=\int\limits_0^{1/T}\langle S_{\bf
k}^\alpha(t+i\lambda)S_{\bf k}^\alpha\rangle\,d\lambda\,,\label{UNO}
\end{equation}
by solving the following integro-differential 
equation~\cite{Hubbard71,Kawasaki76,MCOU}:
\begin{eqnarray}
\frac d{dt}F_{\bf q}(t)=&&-{2\over N}\sum_{\bf k}
(J_{\bf k}-J_{{\bf q}-{\bf k}})(J_{\bf k}-J_{\bf q}) \times \nonumber\\
&&{T\over{\lambda+J_{\bf k}}}\int\limits_0^t F_ {\bf k}(t-t')
F_{{\bf q-k}}(t-t')F_{\bf q}(t'),
\label{mceq}
\end{eqnarray}
where $J_{\bf k}=J\sum_{\bf a}\exp({{\bf k\cdot a}})$, and $\lambda$ is
a temperature-dependent parameter which fixes the static
susceptibility of the system and can be directly related to $\xi(T)$
\cite{MCOU}. 


By means of the spectral theorem the dynamic cross section
${\cal S}_{\bf k}(\omega)$ can be obtained from the Fourier transform
of $F_{\bf k}(t)$. Thus the longitudinal relaxation time is given
by \cite{Riga}:
\begin{equation}
1/{T_1}={\gamma^2\over 2N}
\sum_{\bf k}{\cal S}_{\bf
k}(\omega)[A^2+B^2(\cos(k_xa)+\cos(k_yb))]\,,
\label{DUE}
\end{equation}
where $\gamma=2\pi\cdot42.576\cdot10^2$ s$^{-1}$Gauss$^{-1}$.
The form factors $[A^2+B^2(\cos(k_xa)+\cos(k_yb))]$
have been derived by assuming the hyperfine coupling of $^1$H
nuclei with the two Cu$^{2+}$ nearest neighbors only. The
coupling constants  have been estimated from the rotation
pattern by rotating the crystal around the $a$- axis in a
field of $94$ kGauss. In this way the transferred hyperfine
interaction constant was directly obtained. The dominant dipolar part
was determined  through lattice sums (details
will be published elsewhere \cite{CFTD2}). The  two coupling
constants turned out $A=2.31$ kGauss and $B=1.46$ kGauss.


The analysis in terms of the theoretical models requires the
knowledge of the exchange constant $J$. {\em Ad-hoc} neutron and
light scattering experiments, leading to the best estimate
of $J$ , have been performed in La$_2$CuO$_4$ \cite{APP}.
Analogous methods to evaluate $J$ in CFTD are under way
\cite{Danesi}. At the moment in CFTD we can assume a value around
$J=80\,K$. Estimates based on the velocity of the spin waves
\cite{Clarke} are in the same range, as well as old data
\cite{vecchi}(within 10\%), in agreement with the
dispersion curve  determined by
neutron scattering near the zone-boundary \cite{Danesi}.


The MC scheme requires the static quantities as an external input. For
this purpose the simplest and self-consistent approach is to use those
pertaining to the two-dimensional spherical model\cite{MCOU}. An
alternative method is to rely on the results obtained for $\xi(T)$ by
the PQSCHA to determine the parameter $\lambda$ appearing in
Eq.~(\ref{mceq}), for $T/J\gtrsim0.35$. The two theoretical curves are
reported in Fig.1.  Furthermore we have used the more precise QMC data
for $\xi(T)$ \cite{K} to calculate the relaxation rate at lower
temperatures, down to the value of $T\simeq20\,K$, where three
dimensional ordering effects are revealed by the NMR experimental
findings. The relaxation rate $1/T_1$ should be weakly temperature
dependent in the quantum critical regime \cite{E}, without any
observable minimum around $T/J=0.5$ ($T=40$ K).  Instead the
experimental data do present such a minimum, which is completely
reproduced by our approach. Moreover, in a wide temperature range the
behavior of $\xi(T)$ is also consistent with the renormalized
classical regime \cite{fat2}.  The minimum in $1/T_1$ is related to
the ${\bf k}$ dependence of the hyperfine form factor in
Eq.~(\ref{DUE}), which is strongly peaked at the center of the BZ but
non zero at the AF wave-vector ${\bf k}=(\pi/a,\pi/a)$.  One has to
notice that in the temperature range far from the transition the
spherical model seems to be a good approximation, as shown also by the
temperature-dependence of the correlation length in Fig.2.




The remarkable agreement between MC calculations and the experimental
results, allows us to
derive the correlation length $\xi(T)$ from the
experimental data through an \emph{inversion} of the mode-coupling 
approach. The results are shown in Fig.2
together with the theoretical predictions. The
success of the mode-coupling approach gives further support to the
validity of the scaling hypothesis, and the correlation length can be
extracted with satisfactory accuracy from the relaxation rate, by means
of a procedure which has already yielded reliable quantitative
estimates in La$_2$CuO$_4$ (see Refs.\cite{G,GG}
for details). Indeed, when $\xi(T)\gg a$ classical
scaling arguments for the generalized susceptibility and the decay
constants lead to the following equation \cite{G}:
\begin{eqnarray}
1/{T_1}=\gamma^2\epsilon{S(S+1)\over 3}\left({{\xi}\over{
a}}\right)^{z+2} {\beta^2\sqrt{2\pi}\over\omega_e}\left({a^2 \over
4\pi^2}\right)\times
\cr
\int_{BZ}d{\bf q}
{[A^2+B^2(cos(q_xa-\pi)+cos(q_ya-\pi))]\over
(1+q^2\xi^2)^2}\,
\label{TRE}
\end{eqnarray}
thus establishing a one-to-one correspondence between $1/T_1$ and $\xi(T)$.
In Eq.~(\ref{TRE}) the $2$-d wavevector {\bf q} starts from the BZ boundary
$(\pi/a,\pi/a)$, 
corresponding to the AF ordering wavevector; 
$\omega_e$ is the Heisenberg exchange frequency describing
the fluctuations in the limit of infinite temperature, $z=1$ 
the dynamical scaling exponent, $\epsilon=0.33$ takes into account the
reduction of the amplitude due to quantum fluctuations
\cite{bib311} and $\beta$ is a normalization factor preserving the
sum rule for the amplitude of the collective modes. Also the data
for $\xi(T)$ obtained by means of this procedure 
are reported in Fig. 2. It is apparent that the  values for 
$\xi(T)$  are  close to the ones deduced by mode-coupling approach, the main
differences being at low values of $\xi$, as expected.


From the comparison with the theoretical estimates, the
permanence of the system in the renormalized classical regime, on moving
towards the classical one at highest temperatures, is
shown. No evidence of crossover to the quantum critical regime 
appears, as is clearly shown in the inset of Fig.2.
Finally we point out that although we used a value of $J=80\,K$ in all 
our calculations, the main conclusions
derived regarding the absence of a crossover towards a
quantum critical regime do not depend on $J$.


Summarizing, we have used a single crystal of a proper 2DQHAF which allows one 
to analyze the behavior of  the in-plane correlation 
length $\xi(T)$ over a wide temperature range in the light of several
theoretical models.
$\xi(T)$ has been derived from $^1$H NMR relaxation rates, having verified
also that 
the mode-coupling  theory is able to reproduce successfully the temperature
dependence
of $1/T_1$. It turns out that $\xi(T)$ is well described by QMC, PQSCHA and HTE.
In particular, at variance with other recent theories for 2DQHAF, no
crossover to the 
quantum critical regime occurs, up to temperatures $T/J\simeq 1.5$,
where the classical behavior has already set on. Finally, Eq. 6, which is
based on the validity of the scaling hypothesis, is shown to yield 
reliable estimates of $1/T_1$ even for $\xi\simeq a$.




The authors would like  to acknowledge the scientific correspondence
and discussions with H.M. R{{\o}}nnow and  D.F. McMorrow. We thank also
R.Vaia for discussions and R.Birgeneau and Y.S. Lee  
for exchange of information.





\begin{figure}
\caption{Proton spin-lattice relaxation $1/ T_1$ as a function of temperature. 
  Squares: experimental data;
  dashed-line: theoretical mode-coupling (MC) results using the static
  quantities of the spherical model; full line: MC results using the
  correlation length given by the PQSCHA for $T/J > 0.35$; filled circles:
MC results using the
  correlation length given by QMC simulations for $T/J < 0.35$
{\protect\cite{KimLT97}}.
  }
\end{figure}


\begin{figure}
\caption{Correlation length $\xi$ vs temperature. Squares: experimental data 
  deduced from $1/ T_1$ by inversion of Eq.~({\protect\ref{TRE}}); 
  circles: experimental data
  deduced from $1/ T_1$ by inversion of mode-coupling results 
  leaving $\lambda$ as a free parameter; 
  dashed-line: 
  spherical model; full line:
  PQSCHA; triangles:
  QMC simulations {\protect\cite{KimLT97}}. In the inset we report the
corresponding data
 for $\xi(T)$ derived from Eq.~({\protect\ref{TRE}}) and from the 
spherical model as a function of the
 inverse temperature. According to the predictions of the QNL$\sigma$M 
the QCR should occur for $0.4<T/J<0.8$ 
(i.e. $ 0.016 {\protect\lesssim} 1/T {\protect\lesssim} 0.036 $), 
 with a correspondent linear behavior of $\xi(T)$ vs. $1/T$, at variance
with the experimental
findings. 
}
\end{figure}


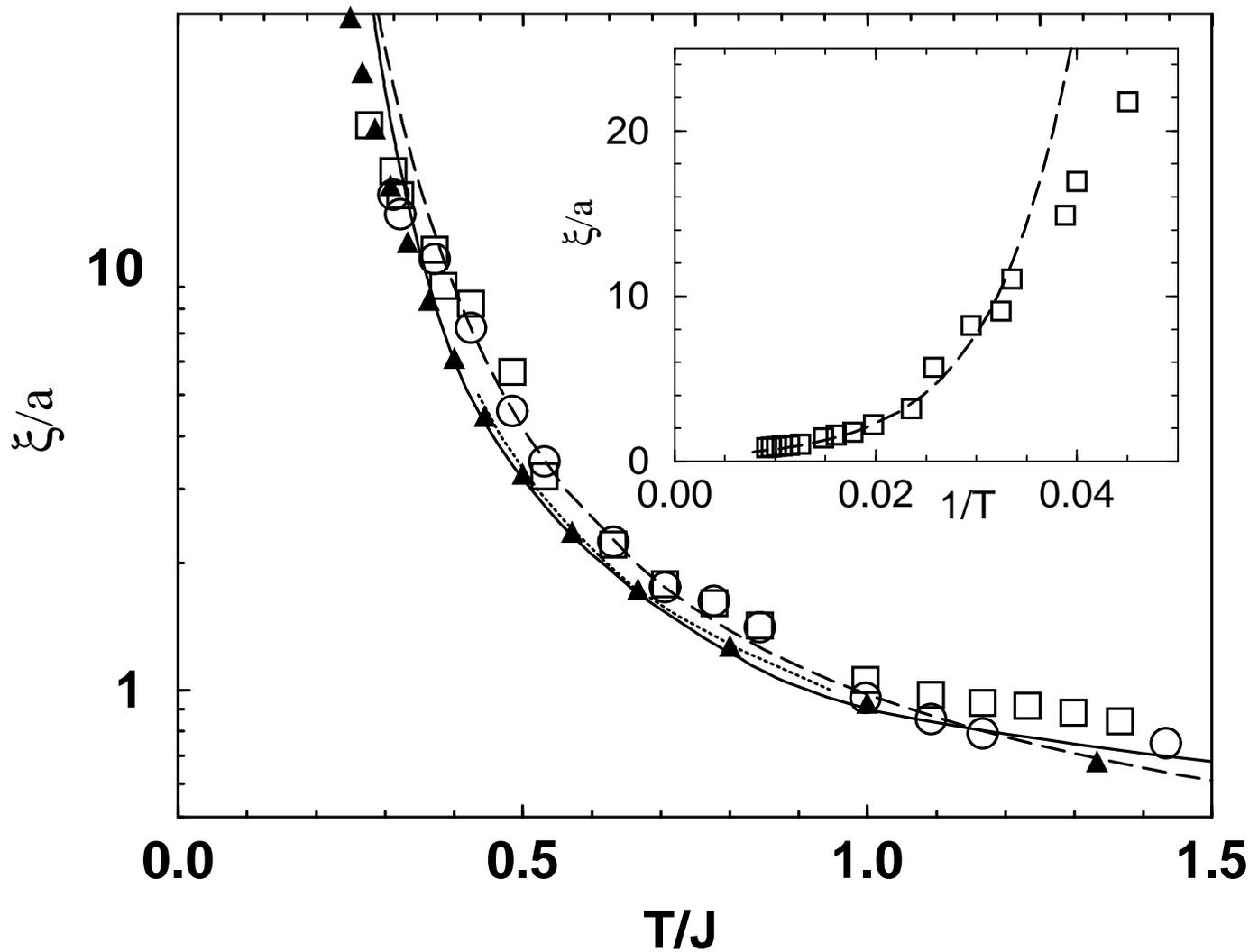
\end{document}